\documentclass[letterpaper, 10 pt, conference]{ieeeconf}  

\IEEEoverridecommandlockouts                              

\usepackage[utf8]{inputenc}
\usepackage[T1]{fontenc}
\usepackage{todonotes}
\usepackage{graphicx}
\usepackage{xfrac}
\usepackage{caption} 
\usepackage{subfig}
\usepackage{float}
\usepackage{amsfonts}

\renewcommand\IEEEkeywordsname{Keywords}

\title{Channel Attention Networks for Robust MR Fingerprinting Matching}

\author{ Refik~Soyak, Ebru Navruz, Eda Ozgu Ersoy, Gastao Cruz, Claudia Prieto, Andrew P. King, Devrim Unay, Ilkay~Oksuz

\thanks{Copyright (c) 2020 IEEE. Personal use of this material is permitted. However, permission to use this material for any other purposes must be obtained from the IEEE by sending a request to pubs-permissions@ieee.org. 

 This work was funded by the Scientific and Technological Research Council of Turkey (TUBITAK) grant number 118C353. This work was supported by an EPSRC programme Grant (EP/P001009/1) and the Wellcome EPSRC Centre for Medical Engineering at the School of Biomedical Engineering and Imaging Sciences, King’s College London (WT 203148/Z/16/Z). The GPU used in this research was generously donated by the NVIDIA Corporation. Correspondence to I. Oksuz.}     

\thanks{R.Soyak, E.Navruz and E.O.Ersoy are with Electrical and Electronics Engineering Department, Izmir University of Economics, Izmir, Turkey.}
\thanks{D.Unay is with Faculty of Engineering, Izmir Democracy University, Izmir, Turkey.}

\thanks{I. Oksuz, Gastao Cruz, Claudia Prieto and Andrew P. King are with School of Biomedical Engineering and Imaging Sciences, King's College London, London, UK.}
\thanks{I. Oksuz is with Computer Engineering Department, Istanbul Technical University, Istanbul, Turkey. E-mail: (oksuzilkay@itu.edu.tr)}}

\begin{document}

\maketitle
\thispagestyle{empty}
\pagestyle{empty}

\begin{abstract}
Magnetic Resonance Fingerprinting (MRF) enables simultaneous mapping of multiple tissue parameters such as T1 and T2 relaxation times. The working principle of MRF relies on varying acquisition parameters pseudo-randomly, so that each tissue generates its unique signal evolution during scanning. Even though MRF provides faster scanning, it has disadvantages such as  erroneous and slow generation of the corresponding parametric maps, which needs to be improved. Moreover, there is a need for explainable architectures for understanding the guiding signals to generate accurate parametric maps. In this paper, we addressed both of these shortcomings by proposing a novel neural network architecture consisting of a channel-wise attention module and a fully convolutional network. The proposed approach, evaluated over 3 simulated MRF signals, reduces error in the reconstruction of tissue parameters by 8.88\% for T1 and 75.44\% for T2 with respect to state-of-the-art methods. Another contribution of this study is a new channel selection method: attention-based channel selection. Furthermore, the effect of patch size and temporal frames of MRF signal on channel reduction are analyzed by employing a channel-wise attention. 

\end{abstract}

\section{Introduction}
\label{intro: 1}
Magnetic Resonance Imaging (MRI) is an essential technique to visualize organs and structures inside the body with applications in basic and clinical sciences. However, although MRI is an adaptable and powerful tool for imaging, it is limited to qualitative imaging that hinders its use in objective evaluation. To overcome this limitation problem, Ma et al.~\cite{Ma2013} introduced Magnetic Resonance Fingerprinting (MRF) as a quantitative MRI solution in 2013. 

The key benefit of MRF is that it provides the opportunity to acquire and quantify different tissue parameters such as the longitudinal relaxation time (T1) and the transverse relaxation time (T2) within a single acquisition, and thus eliminating the need for multiple acquisitions. In MRF, acquisition parameters are varied pseudo-randomly so that each tissue generates a unique signal evolution or fingerprint. Mapping generation is conventionally carried out by applying a dictionary (template) matching algorithm where the acquired signal is matched with the dictionary signal. However, this technique has some shortcomings that need to be addressed such as the computational time required for the dictionary matching algorithm. In dictionary matching algorithm, each acquired signal needs to be compared with the simulated signals resulting in high computational complexity. As the number of combinations in the dictionary increases, reconstruction becomes more expensive with regard to time and storage \cite{Hoppe2019a}. In addition and perhaps more importantly, MRF needs to make a trade-off between accuracy and scanning time, and therefore relies on high under-sampling in k-space. As consequence, dictionary matching may lead to erroneous quantification in the generated maps \cite{Wang2014}.

\section{Related Works}

In this section we provide an overview of the previous studies on dictionary matching in MRF, and the neural network techniques that aim to accelerate this process.

\subsection{Dictionary Matching}

In the original proposal of MRF~\cite{Ma2013}, a dictionary matching algorithm is used to reconstruct the corresponding tissue parameters. Pattern recognition was employed to map the T1 and T2 tissue parameters by matching the fingerprints with a pre-defined dictionary of predicted signal evolutions. Later, different dictionary matching algorithms were proposed to obtain faster and more accurate results. McGivney et al.\cite{McGivney} compressed the dictionary by using singular value decomposition (SVD), enabling faster computation due to the smaller dictionary size. Gomez et al.\cite{Gomez} analyzed the spatiotemporal dictionary and compared it with the temporal MRF dictionary. Besides, some studies suggested to use iterative techniques. For instance, Cline et al.\cite{Cline} proposed a technique called AIR MRF that combines dictionary compression and regularization for accelerated matching. Additionally, Zhao et al.\cite{Zhao} proposed a statistical approach that uses maximum likelihood estimation to predict T1 and T2 tissue parameters accurately. 

\label{method:2}
\begin{figure*}[htb]
    \centering
    \includegraphics[width=\textwidth]{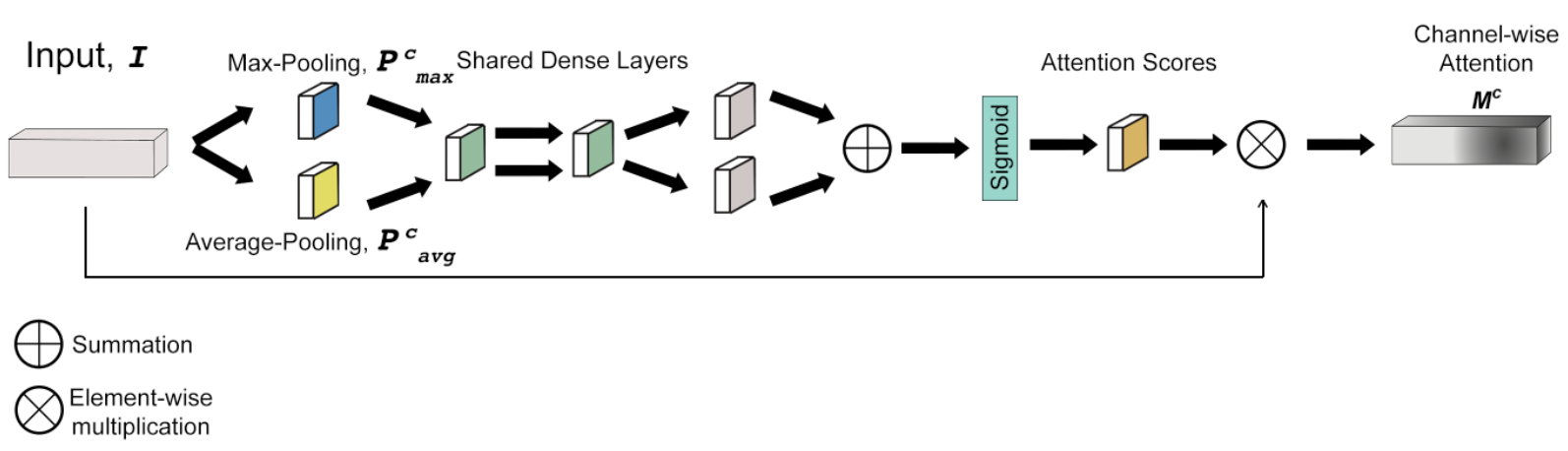}
    \caption{Channel Attention Module Architecture \cite{Woo2018} where weighted input data is produced with attention scores. $P_{max}^{c}$ denotes max-pooled features, $P_{avg}^{c}$ denotes average-pooled features and $M^{c}$ denotes weighted input data.}
    \label{fig:ca_module}
\end{figure*}

\subsection{Neural Network based Techniques}
Successful applications of neural networks for computer vision motivated medical imaging community to use them for accelerating image reconstruction and mapping. Architectures such as Fully Connected Neural Networks (FCN), Convolutional Neural Networks (CNN), and Long Short-Term Memory (LSTM) have been recently investigated for these tasks. Cohen et al.~\cite{cohen2018} proposed a 4-layer fully connected neural network model with two hidden layers by using magnitude images as input. Golbabaee et al.~\cite{Golbabaee} presented a fully connected model with 4 layers that performs dimension reduction with Principal Component Analysis (PCA) at its first layer. Oksuz et al.~\cite{Oksuz2019} used Recurrent Neural Network (RNN) to extract temporal information for the prediction of T1 and T2 values relying on the time series nature of the MRF fingerprint. Likewise, Hoppe et al.~\cite{Hoppe2019} compared CNN and RNN architectures with magnitude and complex-valued inputs, and suggested to use the RNN model with complex-valued MRF signal without using spatial information. 

Balsiger et al.~\cite{Balsiger2019} analyzed both spatial and temporal information to reconstruct tissue parameters using a CNN architecture. Cao et al.~\cite{PengCao} applied a multi-layer perceptron with 4 hidden layers and optimized it to prevent over-fitting. To further increase the speed of training and testing, architectures with pre-processing and pre-training for feature extraction and dimension reduction have been proposed. Since the high dimensionality of the MRF signal requires more computational power to process and creates more redundancy, one of the main focuses of the literature is to reduce signal dimensionality. Thus, a dense network based feature extraction (prior to U-Net) is proposed by Fang et al.~\cite{Fang2019} to extract important information while reducing the number of channels. PCA followed by a Fully Connected Convolutional architecture is used by Chen et al.~\cite{Chen2019}. In order to learn the non-linear relationship between the spatiotemporal MRF image data and multiple quantitative maps, Pirkl et al.~\cite{Pirkl2020DeepLP} proposed to use a CNN architecture combined with relaxation and diffusion-sensitized MRF sequence. 

\par
In this study, we propose a channel attention-based CNN architecture to weight important channels before feeding into the CNN architecture. As our proposed method is attention-based, it uses all channels in a weighted fashion instead of reducing the channels, and therefore eliminates loss of temporal information due to the channel reduction. Furthermore, the use of the channel attention mechanism as proposed allows us in order to examine the relative importance of the channels.\\

Accordingly, there are two major contributions of this work:
\begin{itemize}
    \item To the authors' knowledge, this is the first paper that provides a thorough analysis of attention based methods for magnetic resonance fingerprinting.
    \item A  wide analysis of attention-based channel selections for understanding the significance of each signal in parametric map analysis synthetic.
\end{itemize}

\section{Method}
\label{method:1}

In this section, we will provide the details of our proposed architecture to estimate the MRF parameters.

To empower MRF reconstruction capability, we employed channel-wise attention. Through channel attention, time points are weighted based on attention scores so that the model can understand which time points of the MRF signal are more informative to generate the parametric maps. Channel-wise attention works by aggregating the spatial information of the feature map by using pooling layers to extract inter-channel relationships \cite{Woo2018}. Our suggested channel attention module consists of two pooling layers in parallel, shared dense layers, and a sigmoid activation function. Assume, $P_{max}^{c}$ and $P_{avg}^{c}$ denote max-pooled and avg-pooled features per channel $C$, respectively. These two pooled layers are fed into shared dense layers as shown in Figure~\ref{fig:ca_module}, so that input ($I$) becomes $\mathbb{R}^{1 \times 1 \times C}$. In the first dense layer, channel size is reduced to $\mathbb{R}^{1 \times 1 \times \sfrac{C}{ratio}}$. The $\textit{ratio}$, empirically fixed at 4, is a hyper-parameter that controls the reduction. In the next dense layer, filter size is up-scaled to the original size $\mathbb{R}^{1 \times 1 \times C}$. The outputs of the shared dense layers are summed up and fed into a sigmoid activation function. After the channel attention scores - also known as attention maps - are produced, element-wise multiplication is performed to weight the input $I$. 
The formulation of the channel attention is as follows:

\begin{equation}
\label{eqn: attn_score}
    \alpha = \sigma (Dense(P_{max}^{c}(I)) + Dense(P_{avg}^{c}(I)))
\end{equation}
where $\alpha$ denotes attention scores and $\sigma$ refers to the sigmoid activation function. The dense layers are followed by a Rectified Linear Unit (ReLu) activation function.
\begin{equation}
    \label{eqn: ca}
    M^{c} = I \otimes \alpha
\end{equation}
where $M^{c}$ is the output of attention weighted input $I$. The input channels are weighted by element-wise multiplication ($\otimes$) of input $I$ and attention scores $\alpha$.

\subsection{Proposed Model: $CONV-ICA$}
\label{method:3}

\begin{figure*}[htb]
    \centering
    \includegraphics[width=\textwidth]{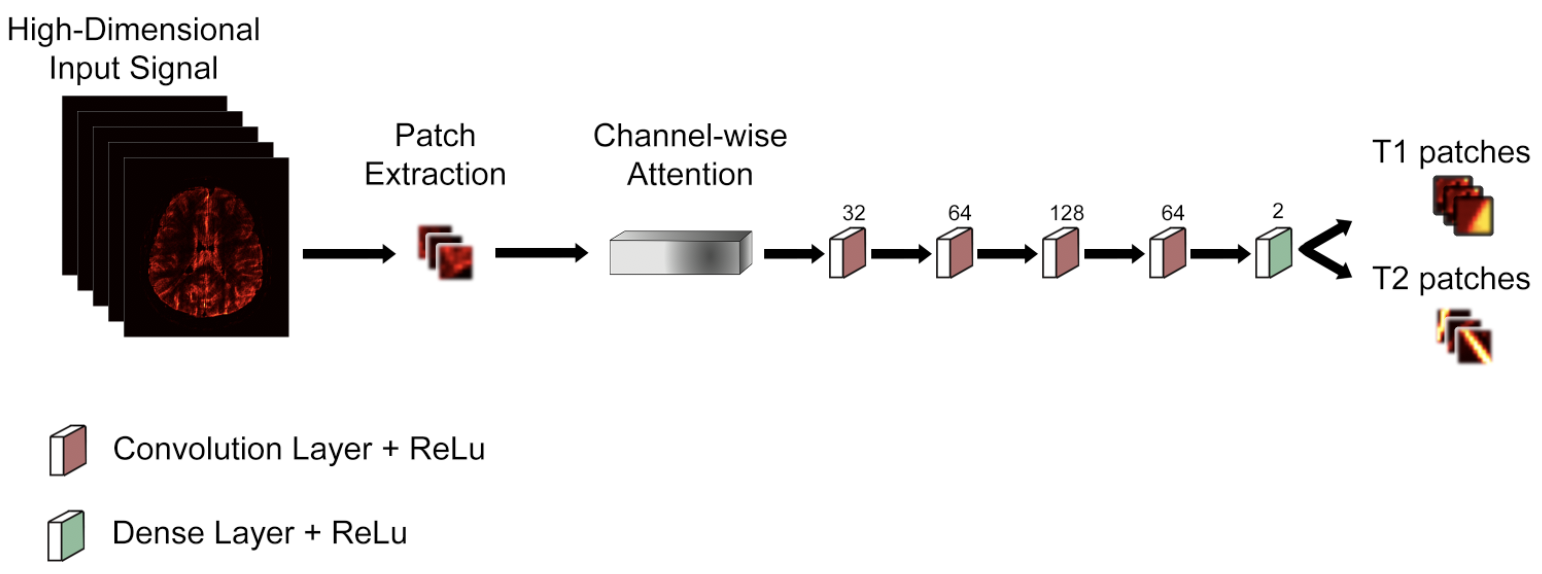}
 
    \caption{Our proposed CONV-ICA model. The algorithm operates on extracted $4 \times 4$ patches.  Filters of convolution layers that follows the channel attention module are determined as $32$, $64$, $138$, $64$ and Dense layer with filter size is $2$ which is the number of tissue parameters is followed by ReLu activation function.}
    \label{fig:proposed_model}
\end{figure*}

In order to calculate the T1 and T2 maps we propose to use a Convolutional network model on Input Channel Attention (CONV-ICA). The current literature for reconstruction of medical images in MRF relies on reducing the number of channels right after the input layer by using a dense layer or convolution layers. However, this reduction causes loss of temporal information. To overcome this problem, this study suggests that use of a channel attention module before reducing channel size (\textit{also known as filter}) as channel attention helps to extract important temporal information by weighting the channels and prevents the loss of that information. As illustrated in Fig.~\ref{fig:proposed_model}, our proposed model consists of a channel attention module and four convolution layers with filter sizes of 32, 64, 128, and 64, respectively. The output layer is a dense layer with the filter size of 2 as it predicts tissue parameters T1 and T2. All layers are followed by the ReLu activation function. The proposed model predicts overlapped patches as shown in Fig.~\ref{fig:proposed_model}.

\subsection{Channel Analysis \& Attention-based Channel Selection}

In MRF, the data consists of many time points (\textit{or channels}), which is 2000 in this study, where some of them might be redundant and therefore may compromise predictions and map reconstructions. Accordingly, analysis of these channels is required to investigate which channels are most informative and important for map reconstruction and to eliminate the redundant ones. For this purpose, Balsiger et al.~\cite{Balsiger2019} zero-filled the channels one-by-one and compared the results to understand the importance of the zero-filled channel. However, such analysis may not be the best way to find the importance of channels as the zero-filled channel could be learned by the network as informative for reconstruction. To eliminate this issue, we propose to use an attention mechanism that will attempt to intrinsically assign attention and score the features of a given feature map by importance. In order to investigate the importance of each channel by benefiting from the advantage of the attention mechanism as explained, we used the channel attention module of Woo et al.~\cite{Woo2018} as shown in Fig.~\ref{fig:ca_module}. 

\begin{figure}[htb]
    \centering
    \includegraphics[width=8cm,height=3cm]{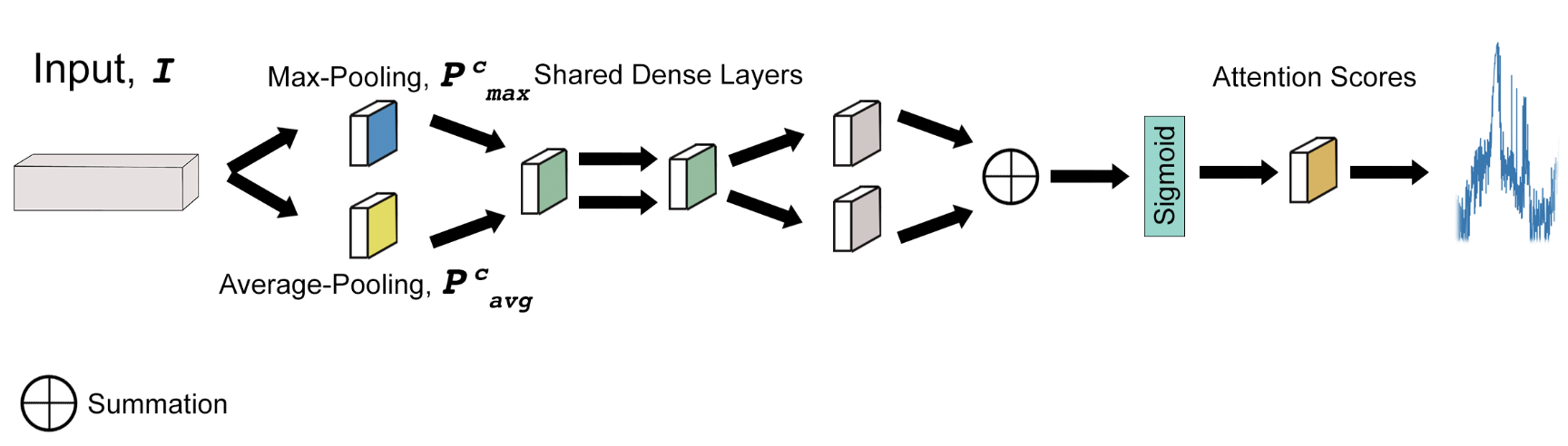}
    \caption{Architecture to produce the attention scores for each channel. Pooling sizes for Max Pooling and Average Pooling are determined as $4\times4$ which is input patch size. Units of dense layers are determined as $500$ by dividing our signal length to the reduction ratio which is decided as $4$ and is up-warded in the next dense layer to input channel length.}
    \label{fig:ca_scores}
\end{figure}

Attention scores are produced after the sigmoid activation function as shown in Fig.~\ref{fig:ca_scores}, and discussed in Section~\ref{method:2}. When attention scores were obtained for each patch, scores of all patches are averaged within each individual channel in order to obtain their distribution over the whole $2000$ channels independently. For attention-based channel selection, \textit{n channels} with the highest attention scores are selected.

\begin{figure}[htb]
    \centering
    \includegraphics[width=8.5cm]{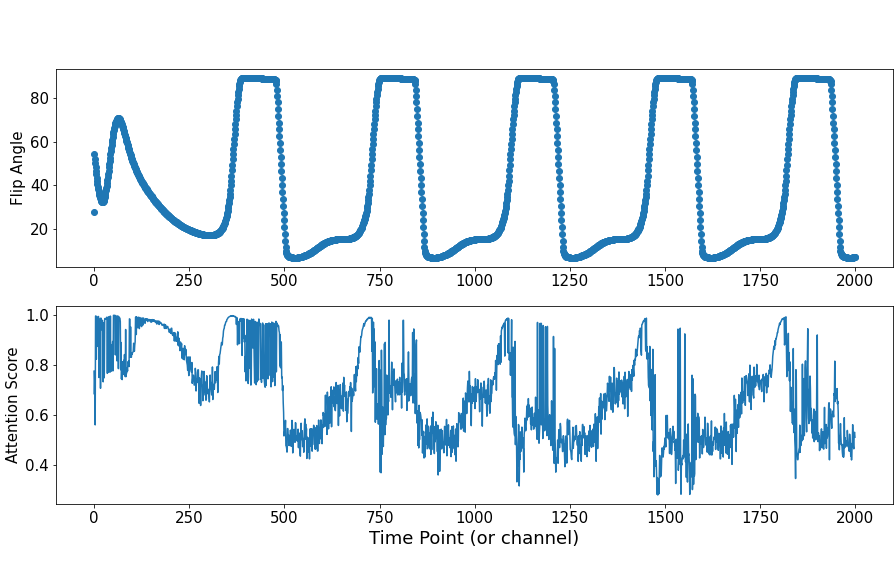}
    \caption{The first plot shows the change of flip angle, where Echo Time (TE) and Repetition Time(TR) were fixed during acquisition, over the 2000 channels. The second plot is the distribution of corresponding attention scores over the 2000 channels obtained from Channel Attention Module, which allows to observe the importance of channels respectively.}
    \label{fig:attn_score_2000}
\end{figure}

The essential channels for tissue parameter estimation could be selected through the channel analysis proposed here, in contrast to the conventional approach where PCA is applied as in~\cite{Chen2019}. We hypothesize that selection of the most important \textit{n channels} by analyzing the distribution of attention scores over all channels (Fig.~\ref{fig:attn_score_2000}) is a better way for channel reduction and elimination of redundant channels.

\section{Experimental Results}

We provide details of our experimental setup and results of our proposed model in comparison to state-of-the-art methods in this section. Methods of comparisons - Balsiger et al. \cite{Balsiger2019}, Chen et al. \cite{Chen2019}, Hoppe et al. \cite{Hoppe2017, Hoppe2019}, and Fang et al. \cite{Fang2019} - are trained and tested by using 3-fold cross validation. As all methods are evaluated over 3 subjects, cross-validation is applied by selecting 1 subject for testing, and the other 2 for training, and continued that process until each subject is selected for testing. Mean absolute error in percentage (Eq.~\ref{eq:MAE}) is used as an evaluation metric.

The proposed model is fed by patch-wise input with size of $4\times4$ and only the magnitude of the signal was used in this study. ADAM~\cite{kingma2014adam} optimization with a learning rate of 15$\times10^{-4}$ was used while training by minimizing the mean squared error. Batch size was experimentally selected as $512$ and the model was trained for $100$ epochs. Early stopping with patience parameter of 15 was used to prevent over-fitting. The Keras~\cite{chollet2015keras} deep learning library with TensorFlow~\cite{tensorflow2015-whitepaper} back-end was used in the Google Colaboratory environment with $25$ GB RAM of GPU.

Equation~\ref{eq:MAE} below defines the mean absolute error in percentage metric, in which $N$ represents the total number of pixels, $\hat{P}_i$ refers to the estimated value and $P_i$ represents the actual value.

\begin{equation}
    \text{Error}= \frac{1}{N} \sum\limits_{i=1}^N \frac{ \left|{P_i}-{\hat{P}_i}\right| } {max(P_{i})}  \times 100
\label{eq:MAE}
\end{equation}

\subsection{Data}
\label{data}

The model is trained and tested on a synthetically generated complex-valued MRF signal. Extended Phase Graph (EPG) \cite{Hennig2004} is used to generate the MRF dictionary for a range of T1 = [0: 2: 500] [500: 5: 1000] [1000: 10: 2000] [2000: 50 : 4000]
ms, and T2 = [0: 1: 100] [100: 2: 500] ms. Repetition Time (TR) is fixed at 4.3ms for utilization of a gradient echo readout.

\subsection{Model Comparison}
 Table~\ref{table:model_comp} shows that proposed approach achieves the best performance for both T1 and T2 parameters as compared to the state-of-the-art models. The results suggest that, using a channel attention mechanism to produce weighted channels leads to a decrease in the reconstruction error for our data set as weighted channels help to avoid losing the important temporal information while reducing the number of channels. Because of under-sampling in k-space during MRF acquisition, the dictionary matching algorithm produces erroneous reconstructions.

\begin{figure*}[ht!]
  \centering
    \includegraphics[width=16cm]{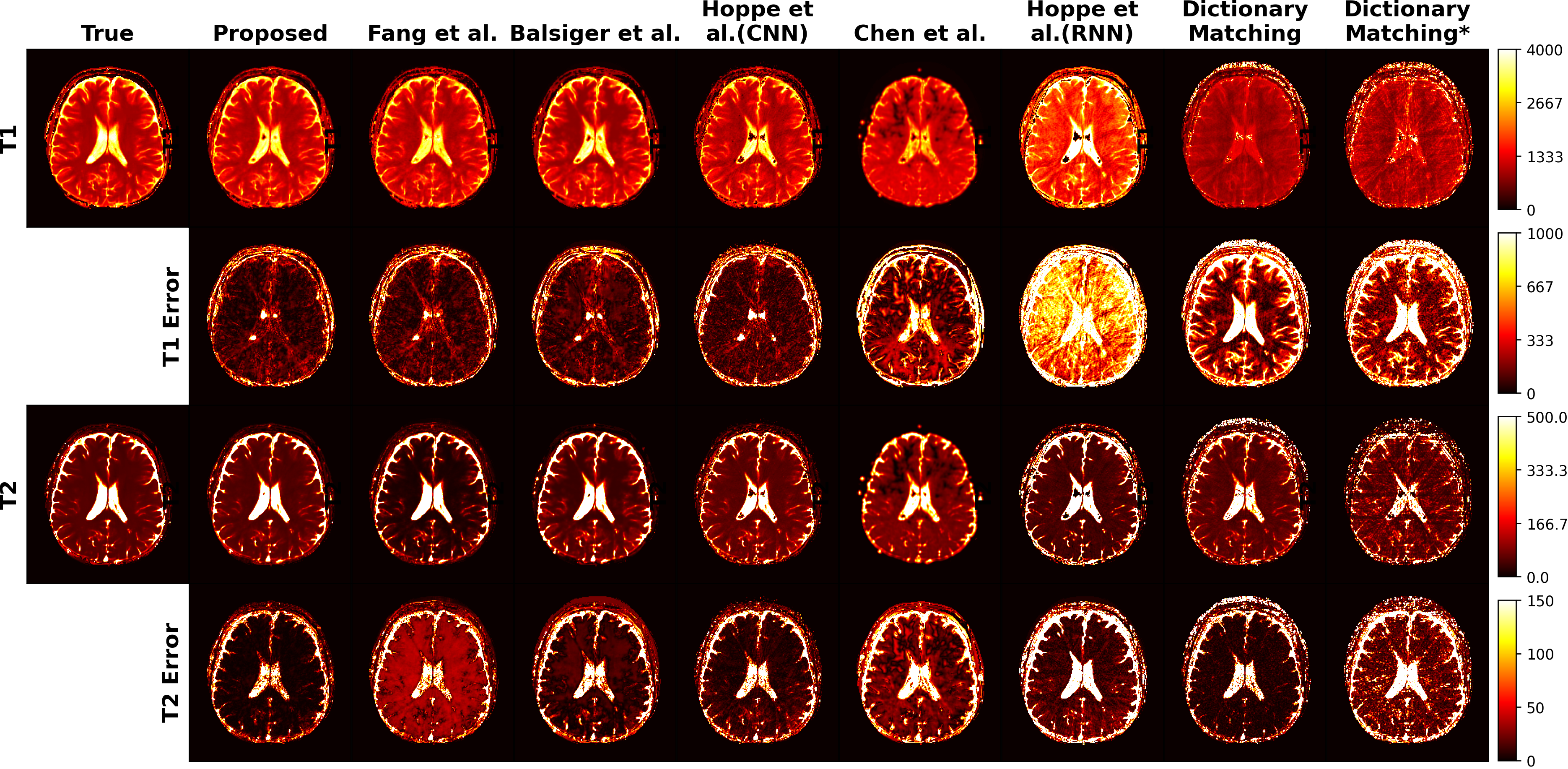}
    \caption{Proposed model to state-of-the-art methods compared qualitatively by testing on a test subject. T1 and T2 errors are the mean absolute differences as shown in Eq.(\ref{eq:MAE}) between reconstructed parameters and true parameters in ms. Backgrounds of reconstructed images are masked to highlight interested region and  errors in Table \ref{table:model_comp} are calculated accordingly. *SVD was used for channel reduction from 2000 to 5 in pre-processing for Dictionary Matching.}
    \label{fig:model_comps}
\end{figure*}

\begin{table}[htp]
\centering
\caption{Comparisons of proposed model with a variety of available techniques where errors are calculated between ground truth images and background-masked reconstructed images as shown in Figure \ref{fig:model_comps} by using Mean Absolute Error (MAE) in percentage. *SVD is used for channel reduction from 2000 to 5 in pre-processing for Dictionary Matching.}
\resizebox{0.45\textwidth}{!}{%
\begin{tabular}{l|cc}
\hline
Models                                 & T1 - MAE(\%) & T2 - MAE(\%)   \\ \hline
Dictionary Matching                    & 4            & 1.01       \\
Dictionary Matching*                   & 4.7           & 1.96        \\
Hoppe et al.\cite{Hoppe2019}~(RNN)       & 5.61          & 1.79    \\
Chen et al.\cite{Chen2019}             & 3.38         & 1.3        \\
Hoppe et al.\cite{Hoppe2017}~(CNN)       & 2.07         & 1.03         \\
Balsiger et al.\cite{Balsiger2019}  & 1.82          & 1    \\
Fang et al.\cite{Fang2019}             & 1.72          & 1.14          \\
Proposed                               & \textbf{1.58} & \textbf{0.57} \\
\hline
\end{tabular}%
}
\label{table:model_comp}
\end{table}

\subsection{Segmentation Comparison}
Table \ref{tab:segmentation-table} expresses comparison of the proposed model and state-of-the-art models for different brain regions, for instance, Skull Stripped, Gray Matter (GM), White Matter (WM), and Cerebrospinal Fluid (CSF) that are extracted by automated segmentation. For segmentation, the Statistical Parametric Mapping (SPM12) \cite{Ashburner2014} is used and errors due to automated segmentation were ignored. Results demonstrate that the proposed method achieved the best reconstruction performance for both parametric maps in all the brain sub-regions. 

\begin{table*}[t]
\centering
\caption{Proposed model and state-of-the-art methods compared quantitatively for segmented brain tissues, such as Skull Stripped, Gray Matter (GM), White Matter (WM), and Cerebrospinal Fluid (CSF). Mean Absolute Errors in percentage~\ref{eq:MAE} are calculated between segmentation of ground truth and reconstructed image for both T1 and T2 tissue parameters.}
\label{tab:segmentation-table}
\resizebox{\textwidth}{!}{%
\begin{tabular}{lcccccccc}
\cline{2-9}
   & \multicolumn{2}{c}{SKULL STRIPPED} & \multicolumn{2}{c}{GM}      & \multicolumn{2}{c}{WM}      & \multicolumn{2}{c}{CSF}     \\ \hline
Models & T1 - MAE(\%)     & T2 - MAE(\%)    & T1 - MAE(\%) & T2 - MAE(\%) & T1 - MAE(\%) & T2 - MAE(\%) & T1 - MAE(\%) & T2 - MAE(\%) \\ \hline
Dictionary Matching  & 2.80 & 0.68 & 0.99 & 0.06 & 0.34 & 0.03 & 2.20 & 0.57 \\
Dictionary Matching* & 2.77 & 1.38 & 0.97 & 0.21 & 0.40 & 0.11 & 2.14 & 1.04 \\
Hoppe et al.\cite{Hoppe2019}~(RNN)           & 3.89 & 1.51 & 2.03 & 0.34 & 1.17 & 0.18 & 2.96 & 1.16 \\
Chen et al.\cite{Chen2019}                 & 1.57 & 0.94 & 0.62 & 0.18 & 0.33 & 0.08 & 1.19 & 0.75 \\
Hoppe et al.\cite{Hoppe2017}~(CNN)       & 0.98 & 0.74 & 0.38 & 0.08 & 0.22 & 0.04 & 0.77 & 0.59 \\
Balsiger et al.\cite{Balsiger2019}        & 0.83 & 0.73 & 0.40 & 0.21 & 0.21 & 0.15 & 0.66 & 0.54 \\
Fang et al.\cite{Fang2019}                 & 0.76 & 0.84 & 0.35 & 0.38 & 0.19 & 0.23 & 0.59 & 0.62 \\
Proposed   & \textbf{0.68} & \textbf{0.37} & \textbf{0.31} & \textbf{0.04} & \textbf{0.15} & \textbf{0.01} & \textbf{0.54} & \textbf{0.29} \\ \hline

\end{tabular}%
}
\end{table*}

\begin{figure*}[htb]
    \centering
    \includegraphics[width=14cm]{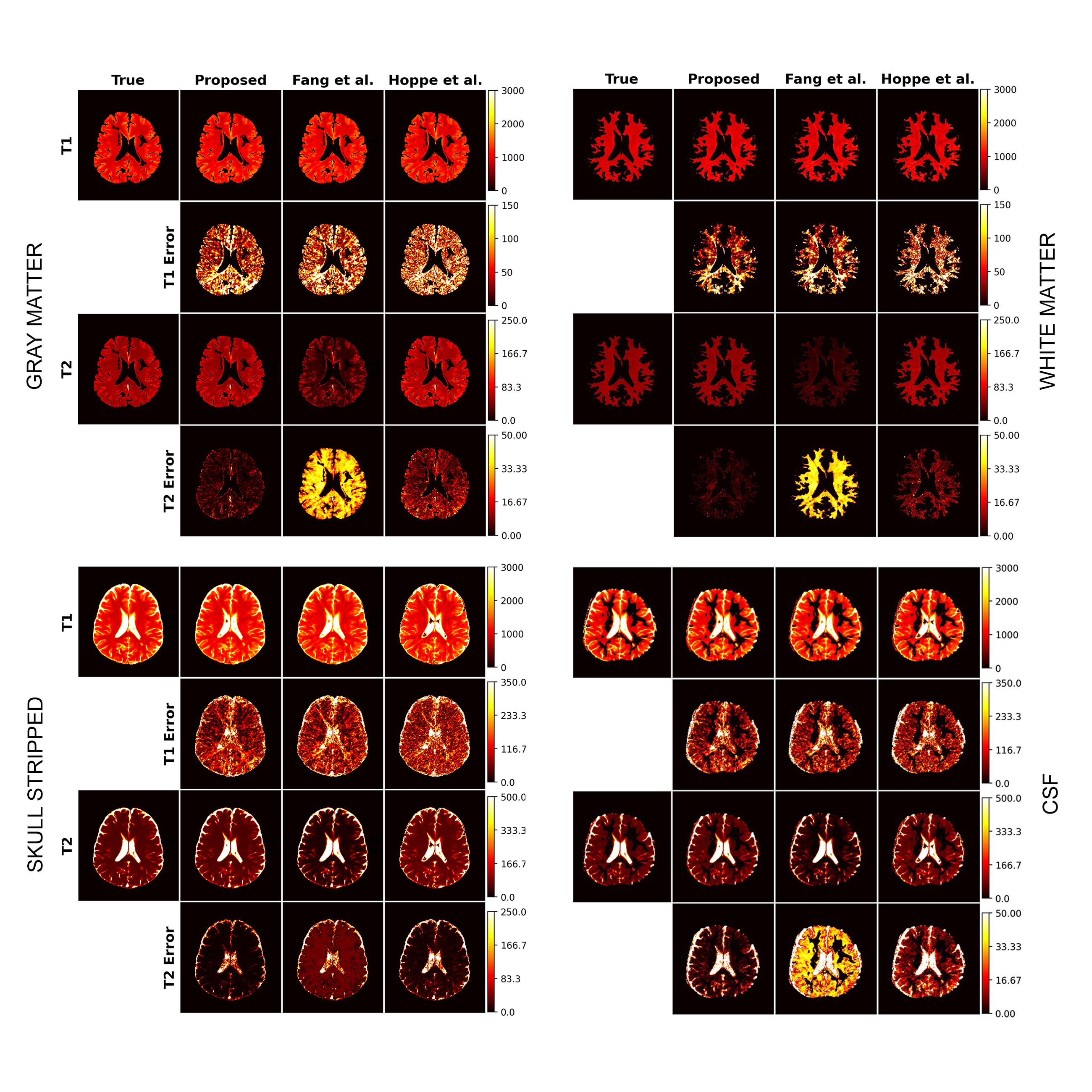}
    \caption{Example results for ground truth (true) image, reconstructions of proposed, Fang et al.\cite{Fang2019} and Hoppe et al.\cite{Hoppe2017} architectures. Four blocks show the reconstruction capabilities for Gray Matter region, White Matter, Cerebrospinal Fluid (CSF) and Skull Stripped respectively. Both T1 and T2 tissue parameters and error maps are visualized for each architecture. }
    \label{fig:segmentation_comparison}
\end{figure*}

\begin{table*}[h]
\centering
\caption{Comparison of number of trainable parameters in million, training time in minutes and prediction time in seconds between the proposed model and various available techniques as well as the dictionary matching method. All methods except Hoppe et al.~\cite{Hoppe2017}, used patch-wise input. Chen et al.~\cite{Chen2019} used PCA and reduced channels to 10 components. *SVD is used for channel reduction from 2000 to 5 in pre-processing for Dictionary Matching.}
\resizebox{0.7\textwidth}{!}{%
\begin{tabular}{l|ccc}
\hline
Model & \multicolumn{1}{l}{\# of Trainable Params (mn)} & \multicolumn{1}{l}{Training Time (min)} & \multicolumn{1}{l}{Prediction Time (sec)} \\ \hline
Dictionary Matching                         & -    & -     & 16392  \\
Dictionary Matching*                        & -    & -     & 13032  \\
Hoppe et al.\cite{Hoppe2019}~(RNN)                & 16.17& 15    & 43.61  \\
Chen et al.\cite{Chen2019}                  & 0.41 & 1.95  & 7.44   \\
Hoppe et al.\cite{Hoppe2017}~(CNN)                & 0.22 & 4     & 1.45   \\
Balsiger et al.\cite{Balsiger2019}          & 9.74 & 61.66 & 145.32 \\
Fang et al.\cite{Fang2019}                  & 1.97 & 58.33 & 18.58  \\
Proposed                           & 2.74 & 30    & 21.81  \\ \hline
\end{tabular}%
}
\end{table*}

The qualitative results presented in Fig.~\ref{fig:model_comps} visually demonstrate the reconstruction performances of the proposed model and the models in comparison for the T1 and T2 parameters. The rows marked as T1 and T2 (1$^{st}$ and 3$^{rd}$ rows, respectively), show the reconstructed tissue parameters by the models. T1 error and T2 error images (2$^{nd}$ and 4$^{th}$ rows, respectively) show the differences between the reconstructed and the true tissue parameters. These qualitative results further support that the proposed model achieves the best performance visually for the reconstruction of T1 and T2 tissue parameters. 

\subsection{Effect of Channel Selection}
The channels are analyzed to investigate the important channels for reconstruction and to eliminate redundant ones. Through the determination of the most important channels, $n$ channels are selected to enhance tissue parameter reconstruction. We proposed a new channel selection method: $attention-based~selection$. To evaluate the performance of the proposed channel selection method empirically, we experimented with two more channel selection schemes, PCA-based and random, and carried out a performance comparison with our proposed method. Table~\ref{table:channel_selection_results} shows MAE errors in percentage (Equation~\ref{eq:MAE}), between the ground truth and reconstructed image for $n$ number of channels selected by the methods: Attention-based, PCA, and Random. Following the channel selection, the proposed model CONV-ICA is used for training. For performance evaluation, 3-fold cross validation is carried out by using the mean absolute error in percentage as a metric. Quantitative results presented in Table~\ref{table:channel_selection_results} show that attention-based channel reduction results in a decrease in the reconstruction error while PCA-based and random selection schemes cause to increase the reconstruction error. The important advantage of attention-based channel selection is reducing the need of resources and run time for algorithms.

\begin{table}[htb]  
\centering
\caption{Comparison of cross validated mean absolute error (MAE) errors in percentage for each channel selection method, attention-based selection, reduction by PCA and random channel selection when $n$ number of channels are selected for both T1 and T2 values.}
\resizebox{0.45\textwidth}{!}{%
\begin{tabular}{l|cc}
\hline
Method / $n$      & T1 - MAE(\%)  & T2 - MAE(\%)  \\ \hline
Attention / 100 & \textbf{2.07} & \textbf{0.79} \\
PCA / 100       & 6.22          & 1.48          \\
Random / 100    & 2.26          & 0.84          \\ \hline
Attention / 200 & \textbf{1.88} & \textbf{0.68} \\
PCA / 200       & 4.49          & 1.71          \\
Random / 200    & 2.08          & 0.69          \\ \hline
Attention / 300 & \textbf{1.93} & \textbf{0.77} \\
PCA / 300       & 4.85          & 1.65         \\
Random / 300    & 2.08          & 0.78          \\ \hline

\end{tabular}
}
\label{table:channel_selection_results}
\end{table}

\begin{figure}[H]
    \centering
    \includegraphics[width=6cm]{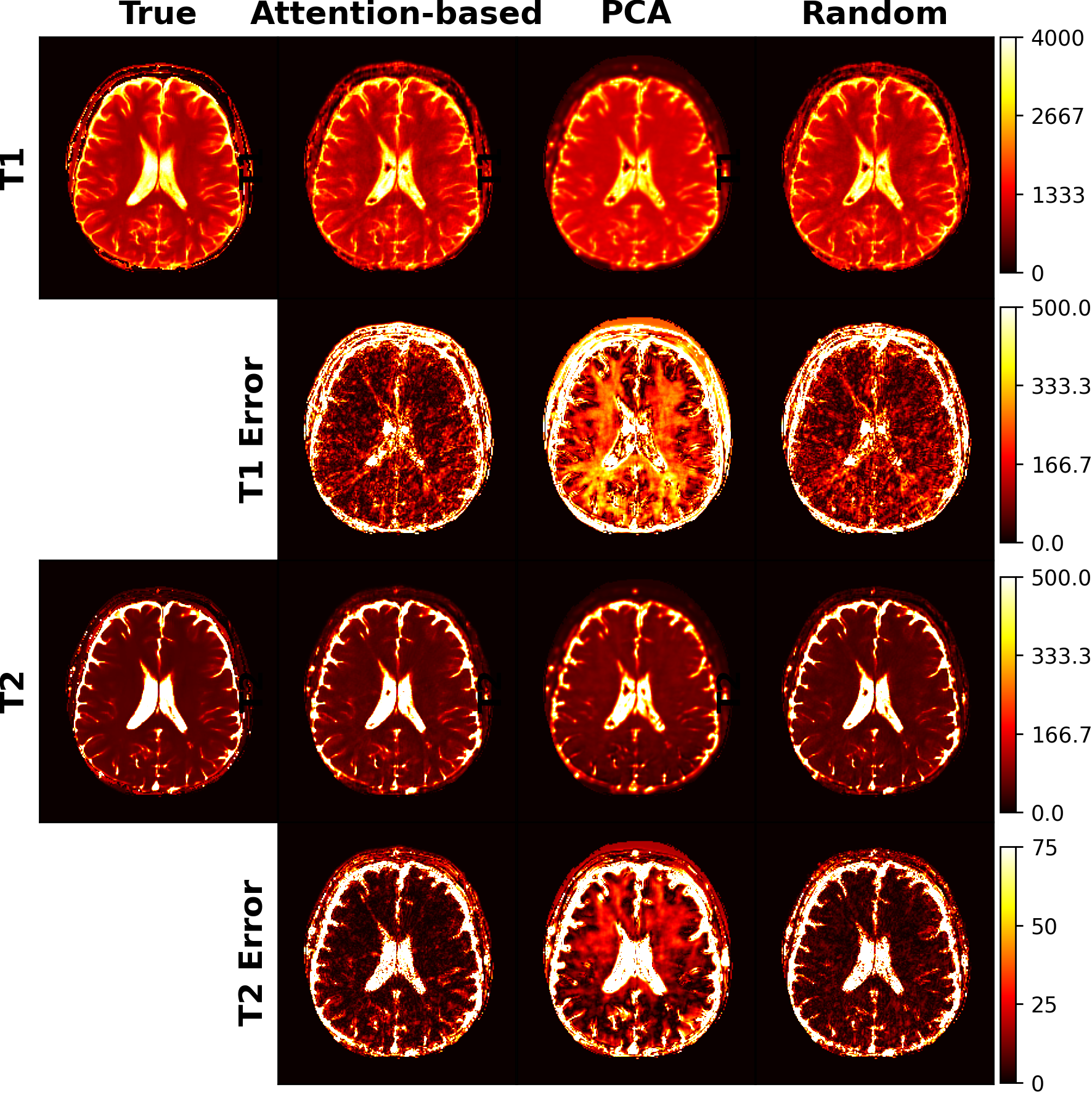}
    \caption{Channel selection methods are compared with true tissue parameters, T1 and T2 respectively, when channel size is reduced from 2000 to 40. The rows marked as T1 and T2 (1$^{st}$ and 3$^{rd}$ rows, respectively), show the reconstructed tissue parameters by the models. T1 error and T2 error images (2$^{nd}$ and 4$^{th}$ rows, respectively) show the differences between the reconstructed and true tissue parameters. }
    \label{fig:channel_selection_comp}
\end{figure}

\subsection{Effect of Patch Size}

In this study, we are utilizing patches to calculate the tissue parameter maps using our architecture. Patch-wise input helps spatial information to be learned by the models.  
In this section, various patch sizes are tested in order to examine the effect of patch size for reconstruction capability. Table~\ref{tab:my-table} shows that the use of 8$\times$8 as the patch size achieves the lowest reconstruction error values for both T1 and T2 tissue parameters. Patch size 12$\times$12 performs the second-best reconstruction for both T1 and T2 value. As a result, it is observed that patch size 8$\times$8 is optimal for the lowest reconstruction error. However, because of insufficient computational resources, patch size was selected as 4$\times$4 in this study. For investigation of the effect of patch size, attention-based channel selection is applied to reduce channel size from 2000 to 40 and patch size is increased accordingly.

\begin{table}[H]
\centering
\caption{Comparison of various patch sizes and reconstruction mean absolute errors (Eq.\ref{eq:MAE}) in percentage respectively. Because of lack of resources, time points were decreased to increase patch size. After attention-based channel selection is performed where channel size is reduced from 2000 to 40, patch sizes are increased and reconstruction errors are observed.}
\label{tab:my-table}
\resizebox{0.45\textwidth}{!}{%
\begin{tabular}{c|cc}
\hline
Patch Size            & T1 - MAE (\%)  & T2 - MAE (\%)  \\ \hline
4$\times$4            & 2.67          & 1.05            \\
8$\times$8.           & \textbf{2.46} & \textbf{0.99}            \\
12$\times$12          & 2.63          & 1.05   \\
16$\times$16          & 2.88          & 1.12            \\
24$\times$24          & 3.02          & 1.19            \\ 
\hline
\end{tabular}%
}
\end{table}

\section{Discussion}

We proposed a new deep learning architecture to address the drawbacks of the dictionary matching algorithm whilst accelerating and improving reconstruction of T1 and T2 values. The proposed method takes advantage of the channel attention mechanism which focuses on important time frames in temporal domain. Therefore, channel attention mechanism at the beginning of the model helps to eliminate the loss of temporal information, while fully convolutional network extracts the spatial information from patch-wise input to reconstruct from MRF signal more accurately than the previously proposed methods. Qualitative and quantitative results demonstrate that the use of the channel attention mechanism at the beginning of the model, where temporal information is mostly lost due to direct channel reduction (e.g. from 2000 to 32), increases reconstruction capability by 8.88\% for T1 value and 75.44\% for T2 value.

In addition to input channel attention, we suggest a new channel selection method: attention-based selection. As MRF data consists of many channels and thus requires high computational resources to process, selection of the most informative channels is needed to decrease the demand of resources such as RAM. Throughout the experiments, it has been observed that approximately 200 channels are sufficient to significantly reduce the need for resources, while maintaining the reconstruction capability. Besides, it helps to accelerate runtime and computational time as data becomes smaller. Additionally and more importantly, during the acquisition of the MRF signal, the parameters are continuously varied to get unique signal evaluations, which may create redundant temporal information~\cite{Oksuz2019}. Therefore, channel selection in MRF data is crucial to keep the most informative and the least redundant temporal frames. Results shows that attention-based selection helps for the best reconstruction among other channel selection methods such as PCA and Random selection for the T1 and T2 tissue parameters. As attention-based channel selection method is deep learning-based, increasing the amount of data will lead to better generalization of channel selection in a more reliable manner, and could outperform random selection significantly in bigger data sets.

In the future, the proposed input channel attention mechanism could be adapted to the U-Net architecture~\cite{ronneberger2015unet} which has shown promising reconstruction capabilities recently~\cite{doi:10.1002/mrm.28136} and other powerful architectures for MRF tissue reconstruction. MRF signal is originally a complex-valued data comprising real and imaginary parts. While the magnitude of the MRF signal is exploited solely in this study (similar to most of the state-of-the-art), in the future it is worth exploring the combined use of real and imaginary parts of the signal as some studies suggest~\cite{Hoppe2019, PatrickVirtue, Barbieri2018CircumventingTC}. Finally, as a future study, we aim to validate our method on larger data sets and other anatomical regions.

In conclusion, this study shows that employing channel reduction at the beginning of the model causes loss of temporal information which is important for the reconstruction of T1 and T2 parameters. To overcome this problem, we proposed deep learning-based input channel attention that can be easily applied to each model. Additionally, we analyzed effect of patch size on reconstruction performance, and furthermore demonstrated quantitatively and qualitatively that the proposed attention-based channel selection achieves the best reconstruction performance.

\bibliographystyle{ieeetr}     
\bibliography{mrf_ref.bib}

\begin{thebibliography}{10}

\bibitem{Ma2013}
D.~Ma, V.~Gulani, N.~Seiberlich, K.~Liu, J.~L. Sunshine, J.~L. Duerk, and M.~A.
  Griswold, ``Magnetic resonance fingerprinting,'' {\em Nature}, vol.~495,
  no.~7440, pp.~187--192, 2013.

\bibitem{Hoppe2019a}
E.~Hoppe, F.~Thamm, G.~K{\"o}rzd{\"o}rfer, C.~Syben, F.~Schirrmacher,
  M.~Nittka, J.~Pfeuffer, H.~Meyer, and A.~K. Maier, ``Rinq fingerprinting:
  Recurrence-informed quantile networks for magnetic resonance
  fingerprinting,'' in {\em MICCAI}, 2019.

\bibitem{Wang2014}
Z.~Wang, Q.~Zhang, J.~Yuan, and X.~Wang, ``Mrf denoising with compressed
  sensing and adaptive filtering,'' in {\em 2014 IEEE 11th International
  Symposium on Biomedical Imaging (ISBI)}, pp.~870--873, IEEE, 2014.

\bibitem{McGivney}
D.~F. {McGivney}, E.~{Pierre}, D.~{Ma}, Y.~{Jiang}, H.~{Saybasili},
  V.~{Gulani}, and M.~A. {Griswold}, ``Svd compression for magnetic resonance
  fingerprinting in the time domain,'' {\em IEEE Transactions on Medical
  Imaging}, vol.~33, no.~12, pp.~2311--2322, 2014.

\bibitem{Gomez}
P.~A. G{\'o}mez, M.~Molina-Romero, C.~Ulas, G.~Bounincontri, J.~I. Sperl, D.~K.
  Jones, M.~I. Menzel, and B.~H. Menze, ``Simultaneous parameter mapping,
  modality synthesis, and anatomical labeling of the brain with mr
  fingerprinting,'' in {\em Medical Image Computing and Computer-Assisted
  Intervention - MICCAI 2016} (S.~Ourselin, L.~Joskowicz, M.~R. Sabuncu,
  G.~Unal, and W.~Wells, eds.), (Cham), pp.~579--586, Springer International
  Publishing, 2016.

\bibitem{Cline}
C.~C. Cline, X.~Chen, B.~Mailhe, Q.~Wang, J.~Pfeuffer, M.~Nittka, M.~A.
  Griswold, P.~Speier, and M.~S. Nadar, ``Air-mrf: Accelerated iterative
  reconstruction for magnetic resonance fingerprinting,'' {\em Magnetic
  resonance imaging}, vol.~41, p.~29—40, September 2017.

\bibitem{Zhao}
B.~{Zhao}, F.~{Lam}, B.~{Bilgic}, H.~{Ye}, and K.~{Setsompop}, ``Maximum
  likelihood reconstruction for magnetic resonance fingerprinting,'' in {\em
  2015 IEEE 12th International Symposium on Biomedical Imaging (ISBI)},
  pp.~905--909, 2015.

\bibitem{Woo2018}
S.~Woo, J.~Park, J.-Y. Lee, and I.-S. Kweon, ``Cbam: Convolutional block
  attention module,'' in {\em ECCV}, 2018.

\bibitem{cohen2018}
O.~Cohen, B.~Zhu, and M.~S. Rosen, ``Mr fingerprinting deep reconstruction
  network (drone),'' {\em Magnetic resonance in medicine}, vol.~80, no.~3,
  pp.~885--894, 2018.

\bibitem{Golbabaee}
M.~Golbabaee, D.~Chen, P.~A. G{\'{o}}mez, M.~I. Menzel, and M.~E. Davies, ``A
  deep learning approach for magnetic resonance fingerprinting,'' {\em CoRR},
  vol.~abs/1809.01749, 2018.

\bibitem{Oksuz2019}
I.~Oksuz, G.~Cruz, J.~Clough, A.~Bustin, N.~Fuin, R.~M. Botnar, C.~Prieto,
  A.~P. King, and J.~A. Schnabel, ``Magnetic resonance fingerprinting using
  recurrent neural networks,'' in {\em 2019 IEEE 16th International Symposium
  on Biomedical Imaging (ISBI 2019)}, pp.~1537--1540, IEEE, 2019.

\bibitem{Hoppe2019}
E.~Hoppe, F.~Thamm, G.~K{\"o}rzd{\"o}rfer, C.~Syben, F.~Schirrmacher,
  M.~Nittka, J.~Pfeuffer, H.~Meyer, and A.~K. Maier, ``Magnetic resonance
  fingerprinting reconstruction using recurrent neural networks,'' {\em Studies
  in health technology and informatics}, vol.~267, pp.~126--133, 2019.

\bibitem{Balsiger2019}
F.~Balsiger, O.~Scheidegger, P.~G. Carlier, B.~Marty, and M.~Reyes, ``On the
  spatial and temporal influence for the reconstruction of magnetic resonance
  fingerprinting,'' in {\em International Conference on Medical Imaging with
  Deep Learning}, pp.~27--38, 2019.

\bibitem{PengCao}
P.~Cao, D.~Cui, V.~Vardhanabhuti, and E.~S. Hui, ``Development of fast deep
  learning quantification for magnetic resonance fingerprinting in vivo,'' {\em
  Magnetic Resonance Imaging}, vol.~70, pp.~81 -- 90, 2020.

\bibitem{Fang2019}
Z.~Fang, Y.~Chen, M.~Liu, L.~Xiang, Q.~Zhang, Q.~Wang, W.~Lin, and D.~Shen,
  ``Deep learning for fast and spatially constrained tissue quantification from
  highly accelerated data in magnetic resonance fingerprinting,'' {\em IEEE
  transactions on medical imaging}, vol.~38, no.~10, pp.~2364--2374, 2019.

\bibitem{Chen2019}
D.~Chen, M.~Golbabaee, P.~A. G{\'o}mez, M.~I. Menzel, and M.~E. Davies, ``A
  fully convolutional network for mr fingerprinting,'' {\em arXiv: Image and
  Video Processing}, 2019.

\bibitem{Pirkl2020DeepLP}
C.~M. Pirkl, P.~A. G{\'o}mez, I.~Lipp, G.~Buonincontri, M.~Molina-Romero,
  A.~Sekuboyina, D.~Waldmannstetter, J.~Dannenberg, S.~Endt, A.~Merola, J.~R.
  Whittaker, V.~Tomassini, M.~Tosetti, D.~K. Jones, B.~H. Menze, and M.~I.
  Menzel, ``Deep learning-based parameter mapping for joint relaxation and
  diffusion tensor mr fingerprinting.,'' {\em arXiv: Medical Physics}, 2020.

\bibitem{Hoppe2017}
E.~Hoppe, G.~K{\"o}rzd{\"o}rfer, T.~W{\"u}rfl, J.~Wetzl, F.~Lugauer,
  J.~Pfeuffer, and A.~K. Maier, ``Deep learning for magnetic resonance
  fingerprinting: A new approach for predicting quantitative parameter values
  from time series.,'' in {\em GMDS}, pp.~202--206, 2017.

\bibitem{kingma2014adam}
D.~P. Kingma and J.~Ba, ``Adam: A method for stochastic optimization,'' in {\em
  ICLR (Poster)}, 2015.

\bibitem{chollet2015keras}
F.~Chollet {\em et~al.}, ``Keras,'' 2015.

\bibitem{tensorflow2015-whitepaper}
M.~Abadi, A.~Agarwal, P.~Barham, E.~Brevdo, Z.~Chen, C.~Citro, G.~S. Corrado,
  A.~Davis, J.~Dean, M.~Devin, S.~Ghemawat, I.~Goodfellow, A.~Harp, G.~Irving,
  M.~Isard, Y.~Jia, R.~Jozefowicz, L.~Kaiser, M.~Kudlur, J.~Levenberg,
  D.~Man\'{e}, R.~Monga, S.~Moore, D.~Murray, C.~Olah, M.~Schuster, J.~Shlens,
  B.~Steiner, I.~Sutskever, K.~Talwar, P.~Tucker, V.~Vanhoucke, V.~Vasudevan,
  F.~Vi\'{e}gas, O.~Vinyals, P.~Warden, M.~Wattenberg, M.~Wicke, Y.~Yu, and
  X.~Zheng, ``{TensorFlow}: Large-scale machine learning on heterogeneous
  systems,'' 2015.
\newblock Software available from tensorflow.org.

\bibitem{Hennig2004}
J.~Hennig, M.~Weigel, and K.~Scheffler, ``Calculation of flip angles for echo
  trains with predefined amplitudes with the extended phase graph
  (epg)-algorithm: principles and applications to hyperecho and traps
  sequences,'' {\em Magnetic Resonance in Medicine: An Official Journal of the
  International Society for Magnetic Resonance in Medicine}, vol.~51, no.~1,
  pp.~68--80, 2004.

\bibitem{Ashburner2014}
J.~Ashburner, G.~Barnes, C.~Chen, J.~Daunizeau, G.~Flandin, K.~Friston,
  S.~Kiebel, J.~Kilner, V.~Litvak, R.~Moran, {\em et~al.}, ``Spm12 manual,''
  {\em Wellcome Trust Centre for Neuroimaging, London, UK}, p.~2464, 2014.

\bibitem{ronneberger2015unet}
O.~Ronneberger, P.~Fischer, and T.~Brox, ``U-net: Convolutional networks for
  biomedical image segmentation,'' in {\em MICCAI}, 2015.

\bibitem{doi:10.1002/mrm.28136}
Z.~Fang, Y.~Chen, S.-C. Hung, X.~Zhang, W.~Lin, and D.~Shen, ``Submillimeter mr
  fingerprinting using deep learning–based tissue quantification,'' {\em
  Magnetic Resonance in Medicine}, vol.~84, no.~2, pp.~579--591, 2020.

\bibitem{PatrickVirtue}
P.~Virtue, S.~Yu, and M.~Lustig, ``Better than real: Complex-valued neural nets
  for mri fingerprinting,'' {\em 2017 IEEE International Conference on Image
  Processing (ICIP)}, pp.~3953--3957, 2017.

\bibitem{Barbieri2018CircumventingTC}
M.~Barbieri, L.~Brizi, E.~Giampieri, F.~Solera, G.~Castellani, C.~Testa, and
  D.~Remondini, ``Circumventing the curse of dimensionality in magnetic
  resonance fingerprinting through a deep learning approach.,'' {\em arXiv:
  Medical Physics}, 2018.

\end{thebibliography}

\end{document}